%Paper: astro-ph/9310012
%From: fadams@mich1.physics.lsa.umich.edu
%Date: Wed, 6 Oct 1993 09:16:54 -0400

\magnification=1000
%\tenpoint
%\singlespace
\hoffset=0.1in
\voffset=0.1in
%\vsize=23.0truecm
%\hsize=16.25truecm
\vsize=22.0truecm
\hsize=5.75truein
%\parskip=0.2truecm

\def\la{\mathrel{\mathpalette\fun <}}
\def\ga{\mathrel{\mathpalette\fun >}}
\def\fun#1#2{\lower3.6pt\vbox{\baselineskip0pt\lineskip.9pt
  \ialign{$\mathsurround=0pt#1\hfil##\hfil$\crcr#2\crcr\sim\crcr}}}
\def\spose#1{\hbox to 0pt{#1\hss}}
\def\lta{\mathrel{\spose{\lower 3pt\hbox{$\mathchar"218$}}
     \raise 2.0pt\hbox{$\mathchar"13C$}}}
\def\gta{\mathrel{\spose{\lower 3pt\hbox{$\mathchar"218$}}
     \raise 2.0pt\hbox{$\mathchar"13E$}}}

%\def\fun{ {\cal F} }
%
%---------------------------------------------------------------------
%
$\,$
\vskip 1.0truein
\centerline{\bf NATURAL INFLATION}
\vskip 0.2truein
\centerline{ {\bf A Talk Presented at the Yamada Conference XXXVII,
Evolution of the Universe and its Observational Quest,
Tokyo, Japan 1993} }
\vskip 0.2truein
\centerline{ {\bf Katherine Freese} }
\vskip 0.1truein
\centerline{\it Physics Department,
University of Michigan}
\centerline{\it Ann Arbor, MI 48109}

\vskip 0.4truein

\centerline{\bf ABSTRACT}
\vskip 0.2truein

A pseudo-Nambu-Goldstone boson, with a
potential of the form $ V(\phi) = \Lambda^4 [1 +
\cos(\phi/f)]$, can naturally give rise to an epoch
of inflation in the early universe (Freese, Frieman, and Olinto 1990).
The potential is naturally flat (as required by microwave background
limits on the amplitude of density fluctuations), without any
fine-tuning.  Successful inflation can be achieved if $f \sim m_{pl}$
and $\Lambda \sim m_{GUT}$. Such mass scales arise in particle physics
models with a large gauge group that becomes strongly interacting at a
scale $\sim \Lambda$, {\it e.g.,} as can happen in the hidden sector
of superstring theories.

The density fluctuation spectrum can be non-scale-invariant, with more
power on large length scales (Adams, Bond, Freese, Frieman, and Olinto
1993).  This enhanced power on large scales may be useful to explain
the otherwise puzzling large-scale clustering of galaxies and clusters
and their flows.  Natural inflation differs from other models with
extra large-scale power in that the contribution of the tensor modes
to microwave background fluctuations is negligible; this difference
should serve as a testable feature of the model.

\vskip .2truein
\centerline {\bf I. INTRODUCTION}
\vskip .2truein
The inflationary universe model was proposed (Guth 1981) to solve
several cosmological puzzles, notably the horizon, flatness, and
monopole problems.  During the inflationary epoch, the energy density
of the universe is dominated by a (nearly constant) false vacuum
energy term $\rho \simeq \rho_{vac}$, and the scale factor $R(t)$ of
the universe expands exponentially: $R(t)= R(t_1) e^{H(t-t_1)}$, where
$H$ = $\dot R /R$ is the Hubble parameter, $H^2 = 8\pi G \rho /3
-k/R^2$ ($ \simeq 8 \pi G \rho_{vac}/3$ during inflation), and $t_1$
is the time at the beginning of inflation.  If the interval of
exponential expansion satisfies $t_{end} - t_1 \ga 65 H^{-1}$, a small
causally connected region of the universe grows to a sufficiently
large size to explain the observed homogeneity and isotropy of the
universe today. In the process, any overdensity of magnetic monopoles
produced at an epoch of Grand Unification is diluted to acceptable
levels.  The predicted GUT abundance of monopoles is $\Omega_{mon}
\simeq 10^{12}$, whereas the energy density of our universe is
observed to be within an order of magnitude of $\Omega =1$; here the
excess monopoles are simply `inflated away' beyond our visible
horizon.  Inflation predicts a geometrically flat universe ($k=0$),
$\Omega \equiv 8 \pi G \rho /3H^2 \rightarrow 1$.

In this talk, I will report on the `natural inflation' proposal.  In
this model, a pseudo-Nambu-Goldstone boson (PNGB) [a particle such as
an axion (Weinberg 1978; Wilczek 1978) or schizon (Hill and Ross
1988)] serves as the field responsible for inflation, the {\it
inflaton}.  In particular, I will focus on three attractive features
of the model:
\hfill\break
1. {\it No `fine-tuning' of the inflaton potential.}
\hfill\break
A successful inflationary model must provide sufficient inflation and
an amplitude of density fluctuations in agreement with observations.
To satisfy these constraints, the potential for the inflaton must be
very flat.  In natural inflation, a PNBG naturally provides a flat
potential, without any fine-tuning of parameters.  I will quantify the
flatness required and will illustrate that it can be obtained
`naturally' with a PNGB.
\hfill\break
2. {\it Density fluctuation spectrum can have extra power
on large scales.}
\hfill\break
The primordial density fluctuation spectrum generated by quantum
fluctuations in the inflaton field is a power law $P(k) \propto k^n$.
In natural inflation a tilted spectrum with $0.6 \leq n \leq 1$ is
obtained. Thus, there can be more power on large length scales than
the scale-invariant $n=1$ Harrison-Zeldovich spectrum.  A smaller $n$
seems favored to match many observations.  In addition, negligible
gravitational wave modes are produced in natural inflation.
\hfill\break
3. {\it Particle physics motivation.}
\hfill\break
Recent work in particle theory has been moving in the direction of
providing the type of particle we need as the inflaton for natural
inflation.

\vskip .2truein
\centerline{\bf II. OUTLINE}
\vskip .2truein

First I briefly sketch how each of these
three features arises. Subsequently I will
provide more detail.

\hfill\break
1. {\it No fine-tuning of the inflaton potential.}
\hfill\break
In `rolling' models of inflation, the accelerated expansion of the
universe takes place while the inflaton `rolls down' the potential
that dominates the energy density of the universe.  The combination of
two constraints, sufficient inflation and amplitude of density
fluctuations in agreement with observations, implies that the
potential must be very flat (Steinhardt and Turner 1984).  To quantify
this statement, we find (Adams, Freese, and Guth 1991) that the ratio
of the height of the potential to the width to the fourth power is
bounded to be quite small, $$\chi = \Delta V /(\Delta \phi)^4 = {{\rm
height} \over {\rm width}^4} \leq 10^{-6}\, .\eqno(1)$$ Most particle
physics models produce $\chi = O(1)$.  For the QCD axion, however,
$\chi \sim 10^{-64}$.  This small number arises because the height and
width are given by two different mass scales (the height by the
logarithmic running of the coupling constant).  The idea behind
natural inflation is to mimic the behavior of the QCD axion or other
PNGBs, only at higher mass scales. The inflaton will then have a flat
potential with a small ratio $\chi$.

The potential for the inflaton will then be $$V(\phi) = \Lambda^4 [1
\pm {\rm cos}(\phi/f)]\, ,\eqno(2)$$ as plotted in figure 1.  The
height and width of the potential are given by two different mass
scales: the height is $2 \Lambda^4$ and the width is $\pi f$.  The
role of these two scales in particle physics will be discussed later.
For the QCD axion the scales are given by $\Lambda_{QCD} \sim 100$ MeV
and $f_{PQ} \sim 10^{12}$ MeV.  For inflation the scales are much
higher: we need $\Lambda \sim m_{GUT} \sim 10^{16}$ GeV and $f \sim
m_{pl}$.  With these scales for the inflaton we have $\chi = (\Lambda
/ f)^4 \sim 10^{-12}$, in agreement with the constraint of eqn. (1).

\vskip .2truein

{\vbox{\vskip 3.5truein}}

\item{}{Figure 1:  Potential of Eq. (5) for natural PNGB inflation;
also, axion potential $V(\phi)$ for temperatures
$T \leq \Lambda$.}
\bigskip

%\hfill\break
2. {\it Spectrum of Density Fluctuations}
\hfill\break
In natural inflation,
the spectrum of density fluctuations
that arises due to quantum fluctuations
in the inflaton field
is $P(k) = |\delta_k|^2 \sim k^n$
where
$n \simeq 1 - {m_{pl}^2 \over 8 \pi f^2}$.
%[Here $\delta_k$ is defined by
%$\delta(x,t) = Sum_k \delta_k e^{i k * x}$,
%where $\rho(x,t) = \bar \rho (t) (1 + \delta (x,t))$
%and the $bar$ denotes the spatial average.]
Many inflationary models give an $n = 1$ scale-invariant spectrum.
Natural inflation gives $0.6 \leq n \leq 1$. The lower limit on $n$
arises because successful reheating in the model requires $f \geq 0.3
m_{pl}$.  Thus natural inflation can provide a tilted spectrum, $n <
1$, with more power on large scales than given by a scale invariant
spectrum.  Ideally we would like to match measurements of microwave
background anisotropies by COBE, large-scale flows, sufficiently early
galaxy formation, and large-scale clustering of galaxies (such as
measured by APM and IRAS).  The first three of these can be
simultaneously matched in a cold dark matter (CDM) scenario for $n
\geq 0.6$ while the fourth prefers $n\leq 0.6$.  Although it is
difficult to simultaneously match all the data, a value of $n < 1$ may
be preferred.

In addition, natural inflation has the feature that it produces
negligible gravitational wave modes.  The same quantum fluctuations
that lead to inflation density perturbations also lead to gravity wave
modes.  Because these are negligible in natural inflation, natural
inflation is less restricted by the combination of observations
mentioned above than are other inflationary models.  Also, the fact
that the tensor modes are negligible is a prediction of the model.
This can be used to discriminate between inflationary scenarios once
we can separate tensor from scalar components in the data (see the
talk by Steinhardt in this volume).

\hfill\break
3. {\it Particle Physics Motivation}
\hfill\break
An important motivation from the point of view of particle physics is
the naturalness of the small number $\chi$ that characterizes the
flatness of the potential.  In addition, recent work in particle
physics provides the type of particle physics we need.  Examples will
be discussed below.
\vskip .125in

\centerline {\bf III. THE MODEL}
\vskip .125in

As mentioned above, to satisfy a combination of constraints on
inflationary models, the potential of the inflation must be very flat,
$\chi \equiv \Delta V/(\Delta \phi)^4
\le {\cal O}(10^{-6} - 10^{-8}) \, ,$
where $\Delta V$ is the change in the potential $V(\phi)$
and $\Delta \phi$ is the change in the field $\phi$
during the slowly rolling portion of the
inflationary epoch.
%(For extended inflation, $\chi \le
%{\cal O}(10^{-15})$ [8].)
Thus, the inflaton must be extremely weakly self-coupled, with
effective quartic self-coupling constant $\lambda_{\phi} < {\cal O}(\chi)$
(in realistic models, $\lambda_{\phi} < 10^{-12}$).

Thus inflation requires a small ratio $\chi^{1/4}$ of two mass scales.
One can take two different attitudes about this small ratio: 1) One
can wait until the heirarchy of mass scales in particle physics is
understood.  We know there is a heirarchy problem as yet unexplained
in particle physics.  Perhaps, in the future, when this is understood,
an explanation for the small ratio of scales in inflation will arise.
2) One can use ideas that exist in particle physics today.  At present
we know of two ways to get small masses ($m \ll m_{pl})$ in particle
physics -- (i) supersymmetry protects small masses, and (ii) Goldstone
bosons.  Here we will take the second attitude and use Goldstone
bosons to look for an explanation for the small ratio of mass scales
required for inflation.  Natural inflation takes advantage of
pseudo-Nambu-Goldstone bosons in particle physics that can naturally
have potentials with two widely disparate mass scales for height and
width.

For the past ten years, people have realized that rolling fields in
inflation require flat potentials and that the parameters of the
potentials must typically be fine-tuned.  Most particle physics models
require $\chi =O(1)$, in contradiction with eqn. (1).  But we know of
a particle with a small ratio of scales: the `invisible' axion has
self-coupling $\lambda_a \simeq [\Lambda_{QCD}/ f_{PQ}]^4 \simeq
10^{-64}$ (Dine, Fischler, and Srednicki 1981; Wise, Georgi, and
Glashow 1981).  Here we use a potential similar to that for axions in
inflation; we obtain `natural' inflation, without any fine-tuning.

The potential we obtain is of the form $V(\phi) = \Lambda^4 [1 +
\cos(\phi/f)], $
%\eqno(3)$$
as in Fig. 1 (the potential takes this form for temperatures $T \leq
\Lambda$).  The height of the potential is $2 \Lambda^4$, and the
width of the potential is $\pi f$.  Thus, the height and the width are
given by two different mass scales, $\Lambda$ and $f$.  As explained
below, $f$ is the scale of spontaneous symmetry breaking of some
global symmetry, and $\Lambda$ is the scale at which a gauge group
becomes strong.  For example, for the QCD axion, $f$ is given by the
Peccei-Quinn scale, $f_{PQ} \sim 10^{15}$ GeV, and $\Lambda$ by the
QCD scale, $\Lambda_{QCD} \sim 100$ MeV.  The ratio of these two
scales to the fourth power is a very small number, $\chi \sim
10^{-64}$.  For inflation, we don't need a ratio quite this small,
but, in order to satisfy various constraints on the model, we will
need the mass scales to be higher.  We can use a particle similar to
the QCD axion (but not the QCD axion itself). Inflation needs $\Lambda
\sim m_{GUT}$ and $f \sim m_{pl}$.
\vskip .125in

\centerline {\bf IIIA. Naturalness}

I will now briefly illustrate how these potentials satisfy the
criterion of naturalness.  I will use the definition of naturalness
proposed by 't Hooft (1979): a small parameter $\alpha$ is natural if,
in the limit $\alpha \rightarrow 0$, the symmetry of the system
increases.  I will show how the axion satisfies this criterion.  [For
references on axion cosmology, see Preskill, Wise, and Wilczek 1983;
Abbott and Sikivie 1983; and Dine and Fischler 1983].  In order to
solve the strong CP problem of QCD, Quinn and Peccei (1977) introduced
a global U(1) symmetry which is broken at a scale $f$, i.e., for $T
\leq f$, the potential of the PQ (Peccei Quinn) field is as
illustrated in Fig. 2a.  For $T \ll f$, the radial modes are frozen
out because they are very massive, and the only remaining degree of
freedom is the angle $\phi/f$ around the bottom of the Mexican
hat-shaped potential.  This angular degree of freedom $\phi$ is the
axion field (Weinberg 1978; Wilczek 1978).

One can plot the value of the potential around the bottom of the
Mexican hat, i.e. $V$ as a function of the angular variable $\phi/f$.
For temperatures $f \geq T \ga \Lambda$, the potential has the same
value for any choice of $\phi$.  This flat potential is plotted in
Fig. 2b.  The Lagrangian for the $\phi$ degree of freedom is just
$L_{\phi} = {1 \over 2} (\partial_{\mu} \phi)^2$, which is invariant
under the transformation $\phi
\rightarrow \phi + const$.  In this sense the U(1)
symmetry is said to be nonlinearly realized (any point around
the bottom of the Mexican hat is equivalent).

For the axion, however, this is not the entire story because of the
chiral anomaly in QCD.  The axion part of the QCD Lagrangian is
$L_\phi = {1 \over 2} (\partial_\mu \phi)^2 +{g^2 \over 32 \pi^2}
{\phi \over f} tr(F\tilde F)$.  At finite temperature, the path
integral from which one extracts the free energy density is dominated
by instanton configurations.

As the temperature drops below $T \leq \Lambda$, instanton effects
turn on, and the bottom of the Mexican hat develops ripples in it.
Not all points in the bottom are equivalent any more.  One can think
of this as placing a block under one side (or under several points) of
the hat, so that the potential as a function of the angular variable
$\phi$ is as in Fig. 1.  This is the cosine potential given in Eq.
(2).  This cosine potential is invariant under the transformation
$\phi \rightarrow \phi + 2 \pi f N$ where $N$ is the number of minima
(ripples) of the potential (in Eq. (2) I have set N=1).  These ripples
around the bottom of the Mexican hat break the nonlinearly realized
symmetry from continuous to discrete.  If one were to set $\Lambda =
0$, then the potential would always retain the flat form of Fig. 2b
for all temperatures, rather than picking up the cosine shape from
instantons.  Thus, taking the limit $\Lambda \rightarrow 0$ restores a
continuous symmetry.  By the definition of naturalness given above,
small $\Lambda$ is therefore natural.  As mentioned before, for the
case of QCD axions, $(\Lambda_{QCD} / f)^4 \sim 10^{-64}$.

\vskip 3.75truein

\item{}{Figure 2:  a) Potential of Peccei Quinn field for $T \leq f$.
The angular
degree of freedom around the bottom is the axion field $\phi$.
b)  Axion potential $V(\phi)$ for temperatures $f \geq T \ga \Lambda.$}
\bigskip

For the inflaton, we used a scalar field with small self-coupling
$\chi$, similar to the case of axions, although the ratio of
parameters does not need to be as small.  Thus, we considered an
axion-like model with scales $\Lambda$ and $f$ as free parameters; we
found that inflation is successful for $f \sim m_{pl}$ and $\Lambda
\sim m_{GUT} \sim 10^{16}$ GeV.  Here, $f$ is the scale at which a
global symmetry is spontaneously broken, and $\Lambda$ is the scale at
which a gauge group becomes strongly interacting.  These mass scales
arise in particle physics models, as described below.

\centerline{\bf IIIB. Constraints on the Potential}

We considered several requirements on inflationary models with
naturally small couplings provided by PNGBs.  These constraints
illustrate why we need the height of the potential to be $\sim
m_{GUT}$ and the width to be $\sim m_{pl}$.

1) {\it Sufficient inflation}

As we will see, sufficient inflation requires $f \geq 0.06 m_{pl}$.

Since the potential as a function of $\phi$ is flat for temperatures
$T \ga \Lambda$, we assumed that the value of the $\phi$ field is
initially laid down at random anywhere in the range $0 \leq \phi/f
\leq 2 \pi$ in different horizon-sized regions.  In other words, at
any one place in the universe, the value of $\phi/f$ can randomly take
on any value between $0$ and $\pi$.  Once the temperature drops to $T
\leq \Lambda$, the cosine potential appears and $\phi$ starts to roll
down the hill.  Only those parts of the universe which start out with
$\phi$ far enough uphill (close enough to the top of the potential)
will get enough inflation to solve the cosmological problems.  We
estimated the success of our scenario by calculating (as a function of
$f$) the probability of $\phi$ being close enough to the top of the
potential to have sufficient inflation.  We estimated this probability
in two different ways: an a priori and an a posteriori probability
described below.  The requirement that sufficient inflation occured
with a probability O(1) drives the value of $f$ to be near the Planck
mass $m_{pl}$.

a) {\it a priori probability:}

First we looked at a time slice of the universe before inflation to
see what fraction of the universe will inflate sufficiently.  Let's
take $\phi_1$ to be the value of the field at some point in space at
the beginning of inflation.  By demanding that there be at least 60
e-foldings of inflation, we found the maximum value $\phi_1^{max}$ of
$\phi_1$ consistent with sufficient inflation.  In other words, we
found how far down the potential the field could be at the beginning
of inflation, and still give rise to enough inflation.  The fraction
of the universe with $\phi \in [0, \phi_1^{max}]$ inflates
sufficiently.  For $\phi_1$ randomly distributed from $0$ to $\pi f$,
the probability of sufficient inflation is then $\phi_1^{max}/ \pi f$.
For $f = (3, 1, {1 \over 2}, \, {\rm and} \, {1 \over \sqrt{24 \pi}})
m_{pl},$ the a priori probability of sufficient inflation is $(0.7,
0.2, 3 \times 10^{-3}, \, {\rm and} \, 3 \times 10^{-41})$
respectively.  Clearly $f$ must be near $m_{pl}$.

b) {\it a posteriori probability}

However, the a priori probability just described is probably overly
cautious and constraining.  Above, we considered the fraction of the
universe before inflation; in fact, the part of the universe that does
inflate ends up much larger than the part that doesn't, and thus
occupies a larger fraction of the universe.  Here we discuss the
probability of sufficient inflation by considering the fraction of the
universe after inflation that is sufficiently homogeneous and
isotropic; to summarize our result, we find that $f \geq 0.06 m_{pl}$
is required for sufficient inflation.

To obtain an ${\it a posteriori }$ probability, we look at the
universe after inflation is finished.  Those regions that started out
closer to the top of the hill inflated a great deal more and thus
occupy a much larger fraction of the universe at the end of inflation.
Here we consider the ${\it a posteriori}$ probability of inflation,
that is, the fraction of the ${\it final}$ volume of the Universe that
inflated sufficiently.
%If one looks at the
%universe after inflation took place
%and asks the question, what fraction
%of the universe now
We calculated the fraction of the volume of the Universe after
inflation which had inflated by at least 60 e-foldings; we found that
the a posteriori probability for inflation is essentially unity for
$f$ greater than the critical value $f_c \sim 0.06 m_{pl}$.

Those values of $f$ which correspond to non-scale-invariant density
fluctuations with $n < 1$ require the field to start relatively close
to the top of the potential.  This is not a problem (does not lead to
fine-tuning) because, in our model, there are regions of the universe
which start out with $\phi$ near the top of the potential, since the
field is randomly distributed between $0$ and $\pi$ to begin with;
those regions with $\phi$ near the top inflate more and grow much
larger than any other portions of the universe.  In fact, for $f \geq
0.06 \, m_{pl}$, those regions which started out with $\phi$
relatively near the top dominate the universe after inflation.  Thus
it is perfectly reasonable (probability O(1)) that our observable
horizon should be within one of these regions.

2) {\it Density Fluctuations.}

Inflationary models generate density fluctuations with amplitude
$$(\delta \rho / \rho)|_{hor} \simeq H^2 / \dot\phi \, , \eqno(3)$$
where the right hand side is evaluated at the time when the
fluctuation crossed outside the horizon during inflation, and
${(\delta \rho /\rho)}|_{hor}$ is the amplitude of the perturbation
when it crosses back inside the horizon after inflation. Fluctuations
on observable scales are produced during the time period (60 - 50)
e-foldings before the end of inflation.  The largest amplitude
perturbations are produced at 60 e-foldings before the end of
inflation, $${\delta \rho \over \rho}\bigg|_{hor}
\simeq {3 \Lambda^2 f \over m_{pl}^3}\left({8\pi
\over 3}\right)^{3/2} {[1+\cos(\phi^{max}_1/f)]^{3/2} \over
\sin(\phi^{max}_1/f)} \, .\eqno(4)$$
Constraints on the anisotropy of the microwave background
(e.g. the recent measurement by COBE) require
$(\delta \rho/\rho)|_{hor} \leq 5 \times 10^{-5}$, i.e.,
$$\eqalignno{\Lambda&\leq 5 \times 10^{15}\,{\rm GeV\ for}\ f=m_{pl}&(5a)\cr
\Lambda&\leq 9 \times 10^{14}\,{\rm GeV\ for}\ f=m_{pl}/2 \ .&(5b)\cr}$$
Thus, to generate the fluctuations responsible for large-scale structure,
$\Lambda$ should be comparable to the GUT scale, and the inflaton mass
$m_{\phi} = \Lambda^2/f \sim 10^{11} - 2\times 10^{12}$ GeV.

In this model, the fluctuations deviate from a scale-invariant
spectrum. For $f \la 3m_{pl}/4$, the amplitude grows with mass scale
$M$ as $(\delta \rho/\rho)|_{hor} \sim M^{m^2_{pl}/48\pi f^2}$. As a
result, the primordial power spectrum (at fixed time) is a power law,
$|\delta_k|^2 \sim k^n$, with spectral index $n \simeq 1 -
(m^2_{pl}/8\pi f^2)$. The extra power on large scales (compared to the
scale-invariant $n = 1$) may have important implications for
large-scale structure.  A detailed discussion is presented in Adams et
al. (1993) and only a brief discussion here.

In figure 3, cold dark matter models with $n = $1, 0.8, 0.6, 0.4,...,
-1 are compared with the angular correlation function $w_{gg}(\theta)$
determined from the APM Galaxy Survey [Maddox et al 1990].  The APM
survey is a two dimensional survey of several million galaxies to a
depth of 600 $h^{-1}$ Mpc.  At large angles, the structure is still in
the linear regime.  The data suggest $0 \lta n \lta 0.7$ is needed for
the CDM model if biasing is linear on large scales.

\vskip 3.75truein

\item{}{Figure 3:
Cold Dark Matter models (with $n = 1$, 0.8, 0.6, 0.4,..., -1)
are compared with the angular
correlation function $w_{gg}(\theta)$
determined from the APM Galaxy Survey.
%scaled to the depth of the Lick catalogue, at which $1^\circ$
%corresponds to a physical scale of $\sim 5 h^{-1} {\rm Mpc}$ (dots).
No nonlinear corrections were applied to the theoretical power
spectra, but for angular scales above $\sim 1^\circ$ and for amplitude
factors $\sigma_8 \lta 1$, the linear approximation is accurate
[Couchman and Bond 1989].  The theoretical curves are in units of
$(b_g \sigma _8)^2$. The straight line gives the angular correlation
that would result if the behavior of the spatial correlation function
observed over distances $r \lta 10 h^{-1}$ Mpc, $\xi \sim r^{-1.8}$,
were extended to large separations.  Vertical hatchmarks indicate the
allowed region once corrections for systematic errors in the
observations are included.  The data therefore suggest $0 \lta n_s
\lta 0.6$ is needed for the CDM model if biasing is linear on large
scales and if $h \geq 0.5$.}
\bigskip

We pay special attention to the prospects of using the enhanced power
to explain the otherwise puzzling large-scale clustering of galaxies
and clusters and their flows. We find that the standard cold dark
matter (CDM) model with $0\lta n_s \lta 0.6$ could in principle
explain these clustering data, such as the APM galaxy angular
correlation function.  However, the microwave background anisotropies
recently detected by COBE (Smoot et al 1992)
imply such low primordial amplitudes for
these CDM models (that is, bias factors $b_8 \gta 2$ for $n_s \lta
0.6$) that galaxy formation would occur too late to be viable and the
large-scale galaxy velocities would be too small. In fact, combining
the COBE results with the requirement of sufficiently early galaxy
formation ($z_{gf} > 2$) leads to the constraint $n_s \gta 0.63$,
which corresponds to $f \gta 0.3 m_{pl}$ for natural inflation
(virtually the same as the sufficient reheating constraint).  A
comparable bound, $n_s \gta 0.72$, arises by combining COBE with the
inferred large-scale flows.  Although no single value of the spectral
index $n_s$ in the standard cold dark matter model universally fits
the data, a value $n_s \leq 1$ may be combined with other variations
of the standard CDM framework to explain the large-scale structure.
For example, if the baryon density is as high as $\Omega_B = 0.1$ or
the Hubble parameter as low as $H_0 = 40$ km/sec/Mpc, then $n_s \sim
0.7$ with CDM would be at least marginally consistent with the
large-scale clustering data, COBE, large-scale velocities, and the
requirement of sufficiently early structure formation.

%Figure 3 is a very rough compendium of various
%observational results. In addition, theoretical
%power spectra $|\delta_k(t_o)|^2 = T^2(k) |\delta_k(t_i)|^2$
%are roughly drawn for $n=1$ and $n=0.7$.
%Here, $T(k)$ is a transfer function which
%maps the power spectrum at the time of formation
%$t_i$ to the power spectrum today $t_o$.
%Values of $n < 1$ seem to be an interesting
%possibility to explain data on large scales.

{\it Gravitational Wave Modes:}

The same quantum fluctuations that give inflaton density fluctuations
$\delta \rho/\rho$ also give gravitational wave perturbations.  The
ratio of the gravitational wave power spectrum $P_{GW}$ to the
adiabatic density perturbation power spectrum $P_{DP}$ at horizon
crossing is $${P_{GW}^{1/2} \over P_{DP}^{1/2}} = {{\sqrt {32 \pi}}
|\dot \phi| \over 3 m_{pl} H}\, .\eqno(6)$$ For natural inflation, we
find that the gravity waves are exponentially suppressed relative to
the adiabatic scalar fluctuations of the inflaton over the observable
large-scale structure waveband.  In particular, for $f \leq m_{pl}$,
this ratio $< 0.04$ for modes with wavelength equal to the current
Hubble radius.

On the other hand, for $n<1$, the gravitational wave modes are
important (Davis et al 1993; Abbott and Wise 1984)
in many other inflationary models including power law
inflation and extended inflation (La and Steinhardt 1989).
Because of these modes, constraints
on the power law index become more restrictive.  For example, the
combination of COBE and demanding that the redshift of earliest galaxy
formation be bigger than 3 requires $n>0.8$.  Thus, the fact that
gravitational wave modes are negligible in natural inflation has two
interesting consequences: i) weaker restrictions on the power law
index and ii) a prediction --- one can use the relative power of
tensor to scalar modes to discriminate among inflationary models.

3) {\it Reheating.}
At the end of the SR regime, the field
$\phi$ begins to oscillate about the minimum of the potential,
and gives rise to particle and entropy production.
The decay of $\phi$ into fermions and gauge bosons reheats
the universe to a temperature (Steinhardt and Turner 1984)
$$T_{RH} = \left({45 \over 4\pi^3 g_*}\right)^{1/4} \min
\left[\sqrt{H(\phi_2) m_{pl}}\ , \sqrt{\Gamma m_{pl}}\right] \, , \eqno(7)$$
where $g_*$ is the number of relativistic degrees of freedom.
On dimensional grounds, the decay rate is
$\Gamma \simeq g^2 {m_\phi}^3 / f^2 = g^2 \Lambda^6 /
f^5$,
where $g$ is an effective coupling constant. (For example, in
the original axion model (Weinberg 1978, Wilczek 1978)
$g \propto \alpha_{EM}$ for two-photon decay,
and $g^2 \propto (m_{\psi}/m_{\phi})^2$ for decays to light
fermions $\psi$.) For $f = m_{pl}$ and $g_* = 10^3$, we find
$T_{RH} = \min [6\times 10^{14}\ {\rm GeV}\ ,  10^8 g\ {\rm GeV}]$.
Since we generally expect $g \la 1$, the reheat temperature will be
$T_{RH} \la 10^8$ GeV, too low for conventional GUT baryogenesis,
but high enough if baryogenesis takes place at the
electroweak scale.  Alternatively,
the baryon asymmetry can be produced
directly during reheating
through baryon-violating decays of $\phi$ or its decay products.
The resulting baryon-to-entropy ratio is
$n_B/s \simeq \epsilon T_{RH}/m_{\phi} \sim
\epsilon g \Lambda/f \sim 10^{-4} \epsilon g$, where $\epsilon$ is the
CP-violating parameter; provided
$\epsilon g \ga 10^{-6}$, the observed
asymmetry can be generated.
Using $\Lambda(f)$ obtained from density fluctuations,
requiring $T_{RH} > 100 $ GeV (the electroweak
scale), we find that we need $f/m_{pl} \geq 0.3$.
This is where the lower limit on the width of the potential
comes from.

\vskip .125in
\centerline{\bf IIIC. Particle Physics Motivation}

An important motivation is the naturalness
of the model. In addition, we have investigated [Adams et al 1993]
particle physics models which already
accomodate particles of the type we need for
natural inflation.
For example, suppose the gauge symmetry
of the effective theory below the scale
$f \sim m_{pl}$ is a product group $G_1 \times G_2$,
where $G_1$ is a standard grand unified
group which spontaneously breaks down to the standard
model at some scale $M_{GUT}$.
In other words, $G_1$ describes the physics of
ordinary quarks and leptons while $G_2$
might describe a `hidden sector'.
Let $G_2$ be an asymptotically
free non-Abelian gauge theory which
becomes strongly interacting at a scale $\Lambda$
comparable to the GUT scale.
Possibilities include making $G_2$ a
technicolor group. A second motivation for a gauge
group which becomes strongly interacting at
the GUT scale comes from superstring theory.
For example, in the original heterotic string
model, $G_1 \times G_2 = E_8 \times E'_8$.
In the effective field theory arising from superstrings,
an important role is played by the complex
scalar field $S = ReS + i \tilde \phi$.  The real part of
this field is the dilaton; the imaginary
part $\tilde \phi$ is the `model-independent axion'
(Witten 1984; Choi and Kim 1985;
Binetruy and Gaillard 1986 also considered
this possibility in the context of
the models of Dine et al 1985).
A particle of this type could play the role of the
inflaton for natural inflation.
When $G_2$ becomes strong at the GUT scale,
the field $S$ gets a potential $V(S) \sim e^{-cS}$.
The imaginary part is $V(\tilde \phi) \propto {\rm cos} \tilde \phi$,
exactly of the right form.
The role of $V(S)$ in particle physics is to
fix the coupling constants of our universe
described by $G_1$ and
%and, preferably, to break
%supersymmetry in our sector
%at a TeV scale, although I believe this has not yet
%been implemented successfully.
%For example, in  the hidden sector of superstring
%theories, if a large non-Abelian group remains
%unbroken, the running gauge coupling can become
%strong at the GUT scale, i.e. $\Lambda \sim m_{GUT}$;
%a hidden sector group that becomes strong at the
%GUT scale
is suggested as a way to break supersymmetry
in a phenomenologically viable way.
%In this case the role of the PNGB inflaton
%might be played, e.g., by the model-independent axion
%(Witten 1984).
% [if SUSY is broken
%in the hidden sector at the GUT scale,
%then the idea is to have it break SUSY
%in our sector at TeV scale and thus
%protect small Higgs masses and alleviate
%the heirarchy problem.]

I would like to briefly comment
on another attractive feature of the model.
Inflation completely solves the flatness
problem if it begins at $T_c = m_{pl} = 10^{19}$
GeV. If it begins later, then for the
case of the closed universe, the universe
recollapses before $T_c$ is ever reached
unless the universe has an
entropy $\bar S \geq (m_{pl}/T_c)^3$.
It is thus very attractive to begin
inflation at the Planck scale. However,
tensor modes (Krauss and White 1993) seem to require
inflation in rolling fields to have
its last 60 e-foldings begin at or
below the GUT scale, $\leq 10^{16}$ GeV.
In natural inflation, we can have an early
inflationary epoch at the Planck
%does this drop temperature too much?
scale as the radial component of
(the equivalent of) the Peccei Quinn
field rolls down its potential after
the PQ symmetry is broken.
Also, spatial gradients in $\phi$ are
efficiently damped
during this brief inflation period
of the radial field (Linde, private communication).
Then,
subsequently, at $T \leq \Lambda \sim m_{GUT}$, the natural inflation
we have described above can take
place for the last 60 e-foldings
of inflation.

\vskip .125in
\centerline{\bf IV. CONCLUSION}

In conclusion, a pseudo Nambu-Goldstone boson, with
a potential [(Eq. 2)] that arises naturally from
particle-physics models, can lead to successful
inflation if the global symmetry-breaking scale
$f \simeq m_{pl}$ and $\Lambda \simeq m_{GUT}$.
Natural inflation is a model
that has i) no fine-tuning: the
potential is naturally flat,
ii) extra power on large scales as indicated
by observations of large-scale clustering
of galaxies and clusters and their flows,
and iii) motivation from already existent particle
physics models.

\vskip 0.5truein

I wish to thank my collaborators, Fred Adams,
Dick Bond,
Josh Frieman, Alan Guth, and Angela Olinto.
I acknowledge support from NSF (Presidential Young
Investigator Fellowship and Grant NSF PHY 9296020),
the Sloan Foundation (Grant 26722
and a Sloan Foundation Fellowship).

\vskip 0.80truein
\centerline{\bf REFERENCES}

\noindent
L. Abbott and P. Sikivie, {\it
Phys. Lett.} {\bf 120 B}, 133 (1983).

\noindent
L. Abbott and M. Wise, {\it Nucl. Phys.} {\bf B244}, 541
(1984).

\noindent
F.C. Adams, J.R Bond, K. Freese, J.A. Frieman,
and A.V. Olinto, {\it Phys. Rev.} D {\bf 47},
426 (1993).

\noindent
F. C. Adams, K. Freese, and A. H. Guth,
{\it Phys. Rev.} D {\bf 43} 965 (1991).

\noindent
P. Binetruy and M. K. Gaillard, {\it Phys. Rev.} D{\bf 34},
3069 (1986).

\noindent
K. Choi and J. E. Kim, {\it Phys. Lett.} {\bf 154B}, 393 (1985).

\noindent
H. Couchman and J.R. Bond, in {\it Large Scale Structure
and Motions in the Universe}, edited by M. Mezetti {\it et al}
(Kluwer, Dordrecht, 1989), p.335.

\noindent
R. L. Davis, H. Hodges, G. F. Smoot, P. J. Steinhardt
and M. S. Turner, preprint (1992).

\noindent
M. Dine, R. Rohm, N. Seiberg, and E. Witten,
{\it Phys. Lett.} {\bf 156 B}, 55 (1985).

\noindent
M. Dine and W. Fischler, {\it Phys. Lett.} {\bf 120
B}, 137 (1983).

\noindent
M. Dine, W. Fischler, and M. Srednicki, {\it Phys. Lett.}
{\bf 104 B}, 199 (1981).

\noindent
K. Freese, J. Frieman, and A. Olinto,
{\it Phys. Rev. Lett} {\bf 65}, 3233 (1990).

\noindent
A. H. Guth, {\it Phys. Rev.} D {\bf 23}, 347 (1981).

%\noindent
%A. H. Guth and S.-Y. Pi, {\it Phys. Rev. Lett.} {\bf 49},
%1110 (1982).

\noindent
C. T. Hill and G.G. Ross, {\it Phys. Lett.} {\bf 203B}, 125
(1988); {\it Nucl. Phys.} {\bf B311}, 253 (1988).

\noindent
L. Krauss and M. White, {\it Phys. Rev. Lett.} {\bf 69}, 869
(1992).

\noindent
D. La and P. J. Steinhardt, {\it Phys. Rev. Lett.}
{\bf 376}, 62 (1989); {\it Phys. Lett.} {\bf 220 B}, 375 (1989).

\noindent
S. J. Maddox, G. Efstathiou, and  W. J. Sutherland,
{\it Mon. Not. R. astr. Soc.}, {\bf 246}, 433 (1990).

\noindent
J. Preskill, M. Wise, and F. Wilczek,
{\it Phys. Lett.} {\bf 120 B}, 127 (1983).

\noindent
H. Quinn and R. Peccei, {\it Phys. Rev. Lett.} {\bf 38}, 1440
(1977).

\noindent
G.F. Smoot et al, {\it Ap. J. Lett.} {\bf 396}, L1 (1992).

\noindent
P. J. Steinhardt and M. S. Turner, {\it Phys. Rev.} D
{\bf 29}, 2162 (1984).

\noindent
G. 't Hooft, in {\it Recent Developments in Gauge Theories},
eds. G. 't Hooft, {\it et al}, (Plenum Press, New York and London, 1979),
p. 135.

\noindent
S. Weinberg, {\it Phys. Rev. Lett.} {\bf 40}, 223 (1978).

\noindent
F. Wilczek, {\it Phys. Rev. Lett.} {\bf 40}, 279 (1978).

\noindent
M. Wise, H. Georgi, and S. L. Glashow, {\it Phys. Rev. Lett.}
{\bf 47}, 402 (1981).

\noindent
E. Witten, {\it Phys. Lett.} {\bf149B}, 351 (1984).

\bye

\vskip 0.8truein
\centerline{\bf Figure Captions}
\noindent
Figure 1:  Potential of Eq. (5) for natural PNGB inflation;
also, axion potential $V(\phi)$ for temperatures
$T \leq \Lambda$.

\noindent
Figure 2:  a) Potential of Peccei Quinn field for $T \leq f$.
The angular
degree of freedom around the bottom is the axion field $\phi$.
b)  Axion potential $V(\phi)$ for temperatures $f \geq T \ga \Lambda.$

\noindent
Figure 3:
Cold Dark Matter models (with $n = 1$, 0.8, 0.6, 0.4,..., -1)
are compared with the angular
correlation function $w_{gg}(\theta)$
determined from the APM Galaxy Survey.
%scaled to the depth of the Lick catalogue, at which $1^\circ$
%corresponds to a physical scale of $\sim 5 h^{-1} {\rm Mpc}$ (dots).
No nonlinear corrections were applied to the theoretical power
spectra, but for angular scales above $\sim 1^\circ$ and for amplitude
factors $\sigma_8 \lta 1$, the linear approximation is accurate
[Couchman and Bond 1989].  The theoretical curves are in units of
$(b_g \sigma _8)^2$. The straight line gives the angular correlation
that would result if the behavior of the spatial correlation function
observed over distances $r \lta 10 h^{-1}$ Mpc, $\xi \sim r^{-1.8}$,
were extended to large separations.  Vertical hatchmarks indicate the
allowed region once corrections for systematic errors in the
observations are included.  The data therefore suggest $0 \lta n_s
\lta 0.6$ is needed for the CDM model if biasing is linear on large
scales and if $h \geq 0.5$.

\bye

\vskip 1.0truein
\centerline{\bf REFERENCES}
\vskip 0.10truein

\item{[1]} A. H. Guth, {\it Phys. Rev.} D {\bf 23}, 347 (1981).

\item{[2]} For a general review of inflation,
see K. Olive, Phys. Repts. {\bf 190}, 307 (1990).

\item{[3]}  A. D. Linde, {\it Phys. Lett.} {\bf 108 B}, 389 (1982);
A. Albrecht and P. J. Steinhardt, {\it Phys. Rev. Lett.}
{\bf 48}, 1220 (1982).

\item{[4]} F. C. Adams, K. Freese, and A. H. Guth,
{\it Phys. Rev.} D{\bf 43}, 965 (1991).

\item{[5]} G. 't Hooft, in {\it Recent Developments in Gauge Theories},
eds. G. 't Hooft, {\it et al}, (Plenum Press, New York and London, 1979),
p. 135.

\item{[6]} H. Quinn and R. Peccei, {\it Phys. Rev. Lett.} {\bf 38}, 1440
(1977); S. Weinberg, {\it Phys. Rev. Lett.} {\bf 40}, 223 (1978);
F. Wilczek, {\it Phys. Rev. Lett.} {\bf 40}, 279 (1978).

\item{[7]} J. E. Kim,
{\it Phys. Rev. Lett.} {\bf 43}, 103 (1979); M. Dine, W. Fischler, and
M. Srednicki, {\it Phys. Lett.} {\bf 104 B}, 199 (1981); M. Wise, H. Georgi,
and S. L. Glashow, {\it Phys. Rev. Lett.} {\bf 47}, 402 (1981).

\item{[8]} K. Freese, J. A. Frieman, and A. V. Olinto, {\it Phys.
Rev. Lett.} {\bf 65}, 3233 (1990). See also J. A. Frieman, in
{\it Trends in Astroparticle Physics}, eds. D. Cline and R. Peccei,
(World Scientific, Singapore, 1992).

\item{[9]} J. P. Derendinger, L. E. Ibanez, H. P. Nilles, {\it Phys. Lett.}
{\bf 155B}, 65 (1985);

\item{[10]}  E. Witten, {\it Phys. Lett.} {\bf 149B}, 351 (1984);

\item{[11]} See, {\it e.g.,} D. S. Salopek,
J. R. Bond, and J. M. Bardeen, {\it Phys. Rev. D} {\bf 40},
1753 (1989)

\item{[12]} A. Zee,  {\it Phys. Rev. Lett.} {\bf 42}, 417
(1979);{\bf 44}, 703 (1980). L. Smolin, {\it Nucl.
Phys.} {\bf B160}, 253 (1979).

\item{[13]} C. T. Hill and G. G. Ross, {\it Phys. Lett.} {\bf 203 B}, 125
(1988); {\it Nucl. Phys.} {\bf B 311}, 253 (1988).
D. Chang, R. N. Mohapatra, and S. Nussinov, {\it Phys. Rev. Lett.} {\bf
55}, 2835 (1985).

\item{[14]} For a review, see J. E. Kim, {\it Phys. Rep.} {\bf 150}, 1 (1987).

\item{[15]} J. A. Frieman, C. T. Hill, and R. Watkins, Fermilab
preprint Fermilab-Pub-91/324-A, to appear in {\it Phys. Rev.} D.

\item{[16]} For a review, see M. Green, J. Schwarz, and E.
Witten, {\it Superstring Theory}, (Cambridge University Press, Cambridge,
1987).

\item{[17]} E. Witten, ref. 10.

\item{[18]} E. Martinec, {\it Phys. Lett.} {\bf 171B}, 189 (1986);
M. Dine and N. Seiberg, {\it Phys. Rev. Lett.} {\bf 57}, 2625 (1986).

\item{[19]} also considered this possibility in the context of
the models of ref.9.

\item{[20]} R. Rohm and E. Witten, {\it Ann. Phys.} {\bf 170}, 454 (1986).

\item{[21]} M. Dine and N. Seiberg, {\it Phys. Lett.}
{\bf 162B}, 299 (1985).

\item{[22]} N. V. Krasnikov, {\it Phys. Lett.} {\bf 193B}, 37 (1987).

\item{[23]} J. A. Casas, Z. Lalak, C. Munoz, and G. G. Ross,
preprint OUTP-90-07P.

\item{[24]} V. Kaplunovsky, L. Dixon, J. Louis, and M. Peskin,
unpublished; L. Dixon, in {\it Proc. of the 15th APS Div. of Particles
and Fields General Meeting}, Houston, TX, Jan. 3-6, 1990.
J. Louis, in {\it Proc. of the  APS Div. of Particles and
Fields General Meeting}, Vancouver, Canada, Aug. 18-22, 1991.

\item{[25]} S. Shenker, Rutgers preprint RU-90-47 (1990).

\item{[26]} B. Ovrut and S. Thomas, {\it Phys. Lett.} {\bf 267B},
227 (1991); preprint UPR-0455T.

\item{[27]} J. Kim and K. Lee, {\it Phys. Rev. Lett.} {\bf 63}, 20
(1989);

\item{[28]} M. Kamionkowski and J. March-Russell,
preprint IASSNS-HEP-92-9  (1992);
R. Holman, S. Hsu, E. Kolb, R. Watkins, and L. Widrow, preprint
NSF-ITP-92-06. S. Barr and D. Seckel, Bartol preprint.

\item{[29]} A. D. Linde, {\it Phys. Lett.} {\bf 129 B}, 177 (1983).

\item{[30]} T. W. B. Kibble, {\it J. Phys.} {\bf A 9}, 1387 (1976).

\item{[31]} D. Goldwirth, {\it Phys. Lett.} {\bf 243B}, 41 (1990).

\item{[32]} L. Knox and A. V. Olinto, unpublished.

\item{[33]} A. H. Guth and S.-Y. Pi, {\it Phys. Rev. Lett.} {\bf 49},
1110 (1982).

\item{[34]} S. W. Hawking, {\it Phys. Lett.} {\bf 115B}, 295 (1982).

\item{[35]} A. A. Starobinskii, {\it Phys. Lett.} {\bf 117B}, 175 (1982).

\item{[36]} J. Bardeen, P. Steinhardt, and M. S. Turner,
{\it Phys. Rev.} D {\bf 28}, 679 (1983).

\item{[37]} G. F. Smoot, C. L. Bennett, A.  Kogut, E. L. Wright,
J. Aymon, N. W. Boggess, E. S. Cheng,  G. De Amici, S. Gulkis,
M. G. Hauser, G. Hinshaw, C. Lineweaver, K. Loewenstein,
P. D. Jackson, M. Jansen, E. Kaita,  T. Kelsall, P. Keegstra,
P. Lubin, J. Mather, S. S. Meyer,  S. H. Moseley,  T.  Murdock, L. Tokke,
R. F. Silverberg, L. Tenorio,  R. Weiss and D. T. Wilkinson,
{\it Ap.\ J.\ Lett.}, in press  (1992).

\item{[38]}  N. Vittorio, S. Mattarese, and F. Lucchin,
{\it Ap. J.} {\bf 328}, 69 (1988).

\item{[39]} A. Liddle, D. H. Lyth, and W. Sutherland, {\it Phys. Lett.}
{\bf 279B}, 244 (1992).

\item{[40]}  L. A. Kofman and A. D. Linde,  {\it Nucl. Phys.} {\bf B282}, 555
(1987); L. A. Kofman and D. Yu. Pogosyan, {\it Phys. Lett.} {\bf 214B}, 508
(1988).

\item{[41]} E. W. Kolb, D. S. Salopek, and M. S. Turner, {\it Phys. Rev.}
D{\bf 42}, 3925 (1990).

\item{[42]}  J. M. Bardeen, J. R. Bond, N. Kaiser, and A. S. Szalay,
{\it Ap. J.}, {\bf 304}, 15 (1986), BBKS.

\item{[43]}
J. R. Bond   and A. S. Szalay, {\it Ap. J.} {\bf 274}, 433  (1983).

\item{[44]} J. R. Bond and H. Couchman,
in {\it General Relativity and Astrophysics}, Proc. Second
Canadian Conference on General Relativity, p. 385 (1987);
H. Couchman  and J. R. Bond, in: Large Scale Structure and
Motions in the Universe, eds. M.  Mezzetti \etal  (Dordrecht: Kluwer),
p.335  (1989); J. R. Bond and H. Couchman,
in preparation (1992).

\item{[45]}
J. M. Bardeen, J. R. Bond,  and G. Efstathiou,
{\it Ap. J.}, {\bf 321}, 18 (1987).

\item{[46]}

\item{[47]}
J. R. Bond, in {\it Highlights in Astronomy, volume 9},
Proceedings of IAU Joint Discussion IV, Buenos Aires General Assembly,
ed. J. Bergeron,
{\it Structure Constraints from Large Angle CMB
Anisotropies}  (1991).

\item{[48]} N.  Bahcall and R. Soneira,  {\it Ap. J.}, {\bf 270}, 70
(1983).

\item{[49]} G. B. Dalton, G. Efstathiou, S. J. Maddox  and  W. J.
Sutherland, {\it Ap. J.}, in press (1992);
R. C. Nichol, C. A. Collins, L. Guzzo, S. L. Lumsden,
{\it Mon. Not. R. astr. Soc.}, {\bf 255}, 21P (1992).

\item{[50]}
J. R. Bond and G. Efstathiou, {\it Mon. Not. R. astr. Soc.}, {\bf
226},  655 (1987).

\item{[51]}
E. Bertschinger, A. Dekel, S. M. Faber, A. Dressler and
D. Burstein,
{\it Ap. J.}, {\bf 364}, 370 (1990).

\item{[52]}
G. Efstathiou, R. S. Ellis  and B. A. Peterson, {\it Mon. Not. R.
astr. Soc.} {\bf
232}, 431  (1988).

\item{[53]}
W. H. Press  and P. Schechter, {\it Ap. J. } {\bf 187}, 425 (1974).

\item{[54]} J. R. Bond, in {\it Frontiers in Physics --- From Colliders to
Cosmology}, Proc. Lake Louise Winter Institute, p. 182, ed.
A. Astbury, B.A. Campbell, W. Israel, A.N. Kamal and F.C. Khanna,
Singapore: World Scientific (1989).

\item{[55]}
J. R. Bond and S. T. Myers, in {\it Trends in Astroparticle Physics},
eds. D. Cline \& R. Peccei, World Scientific, Singapore, p262 (1992);
in Proceedings of the Third Teton Summer School on {\it Evolution of
Galaxies and Their Environment}, July 5--10, 1992, ed. M. Shull (1992).

\item{[56]} J. R. Bond, S. Cole, G. Efstathiou and N. Kaiser,
{\it Ap. J. } {\bf 379}, 440 (1991).

\item{[57]}
B. Fort,  \etal preprint (1992).

\item{[58]}
K. Arnaud,  \etal preprint (1992).

\item{[59]}
C. S. Frenk, S. D. M. White, G. Efstathiou  and M. Davis,
{\it Ap. J.}, {\bf 351}, 10 (1990).

\item{[60]}
A. S. Dekel and J. Silk,  {\it Ap. J. } {\bf 303}, 39 (1986).

\item{[61]}
M. Davis, G. Efstathiou, C. S. Frenk, and S. D. M. White,
{\it Ap.J.} {\bf 292} (1985).

\item{[62]}
H. M. P. Couchman and  R. G. Carlberg, {\it Ap. J.},
{\bf 389}, 453 (1992).

\item{[63]}
G. Smoot,  \etal {\it Ap. J. Lett.} {\bf 371},  1 (1991).

\item{[64]}
P. Meinhold  and P. Lubin, Ap. J. Lett. 370, 11  (1991).

\item{[65]}
N. Kaiser, G. Efstathiou, R. S. Ellis, C. S. Frenk, A. Lawrence,
M. Rowan-Robinson and W. Saunders,  {\it Mon. Not. R. astr.
Soc.}, {\bf 252}, 1 (1991).

\item{[66]} K. B. Fisher, M. Davis, M. A. Strauss, A. Yahil and J. P.
Huchra, {\it Ap. J.}, in press (1992); K. B. Fisher, Ph. D.
thesis, U. C. Berkeley (1992).

\item{[67]} Y. Suto, N. Gouda, and N. Sugiyama, {\it Ap. J. Suppl.}
{\bf 74}, 665 (1990).

\item{[68]} R. Cen, N. Y. Gnedin, L. A. Kofman and J. P. Ostriker,
preprint (1992).

\item{[72]} A. Dolgov and J. Silk, preprint (1992); L. Krauss and M.
White, preprint (1992); D. S. Salopek, preprint (1992).

\item{[73]} D. La and P. J. Steinhardt, {\it Phys. Rev. Lett.} {\bf
62}, 376 (1989); E. Weinberg, {\it Phys. Rev.} {\bf D40}, 3950 (1989);
D. La, P. J. Steinhardt and E. W. Bertschinger, {\it Phys. Lett.} {\bf
B231}, 231 (1989).

\item{[74]} A. R. Liddle and D. Wands, {\it Mon. Not. R. astr. Soc.}
{\bf 253}, 637 (1991).

\bye